# Stateless and Rule-Based Verification
# For Compliance Checking Applications

Mohammad Reza Besharati

PhD Candidate, Computer Engineering Department, Sharif University of Technology, Tehran, Iran, Corresponding Author, besharati@ce.sharif.edu

Mohammad Izadi

PhD, Associate Professor, Computer Engineering Department, Sharif University of Technology, Tehran, Iran, izadi@sharif.edu

Ehsaneddin Asgari

PhD, Applied Science and Technology, University of California, Berkeley, CA 94720, USA, asgari@berkeley.edu[1]

**Context**: Underlying computational model has an important role in any computation. The state and transition (such as in automata) and rule and value (such as in Lisp and logic programming) are two comparable and counterpart computational models. Both of deductive and model checking verification techniques are relying on a notion of state and as a result, their underlying computational models are state dependent. Some verification problems (such as compliance checking by which an under compliance system is verified against some regulations and rules) have not a strong notion of state nor transition. Behalf of it, these systems have a strong notion of value symbols and declarative rules defined on them. **Objective:** SARV (Stateless And Rule-Based Verification) is a verification framework that designed to simplify the overall process of verification for stateless and rule-based verification problems (e.g. compliance checking). **Method:** In this paper, a formal logic-based framework for creating intelligent compliance checking systems is presented. We define and introduce this framework, report a case study and present results of an experiment on it. **Results**: The case study is about protocol compliance checking for smart cities. Using this solution, a Rescue Scenario use case and its compliance checking are sketched and modeled. An automation engine for and a compliance solution with SARV are introduced. Based on 300 data experiments, the SARV-based compliance solution outperforms famous machine learning methods on a 3125-records software quality dataset. **Conclusions:** The proposed SARV method could be beneficial for stateless verification in compliance checking applications.

---

[1] Currently NLP Expert Center, Volkswagen Group Data:Lab, Munich, Germany

# 1 INTRODUCTION

Every organization or system must be committed to complying with the laws, standards and requirements of its field of activity. The need to adhere to rules and regulations includes any organization (including individual, team, body, department, organization and wider organization) and in general, any system, especially business, social, content, software, other systems.

In general, it can be stated that both compliance to internal regulations and rules and regulations set by governments, regulatory agencies and standardization bodies from outside the organization are important concerns of the organization and businesses today. Therefore, compliance management and related technologies are one of the important components of E-Governance and E-Organizations.

Intelligence cannot be accomplished without paying attention to processes. Many intelligent systems have to deal with process concepts (for example, intelligent systems supporting the healing process and recovery in E-Health). Therefore, the ability to process processes, at the semantic level, is essential for intelligence in many areas. Compliance checking allows semantic evaluation of processes to be done computationally, and is thus required for intelligence in many areas.

This paper presents the SARV framework for creating intelligent compliance checking systems using formal semantics. It is expected that the application of this framework will facilitate the creation of intelligent compliance checking systems.

## 1.1 Proposed formal semantics: Semantic Logic

The proposed formal semantics is a logic. This logic, which we call the Semantic Logic [1], is based on a specific interpretation of intuitionism. Hence, it belongs to the family of Intuitionistic Logics. In the following, an introduction to intention in logic and intuitionistic logics is presented, and then the semantics in the proposed logic - that is, Semantic Logic - is presented.

*1.1.1 Semantics in Logic and Intuitionistic Logics*

Every logic is made of well-structured formulas and the intentions of these formulas (the semantic of these terms). Hence, the meta-model [2] presented in Figure 1 can be considered for logic (as typical and general) [3].

In this meta-model we see that logic consists of two components, "well-constructed formulas" and "intention of formulas", which represent the components of Syntax and Semantic, respectively.

Two important examples of the intention of formulas are "the value of expressions and formulas" and their "valuation rules" [4]. On the other hand, for well-made formulas, the rules for making formulas are used. These rules are defined on a set of symbols called the "alphabet of formulas". The definition of a logic is, in many cases, done by presenting and defining the components shown in Figure 1.

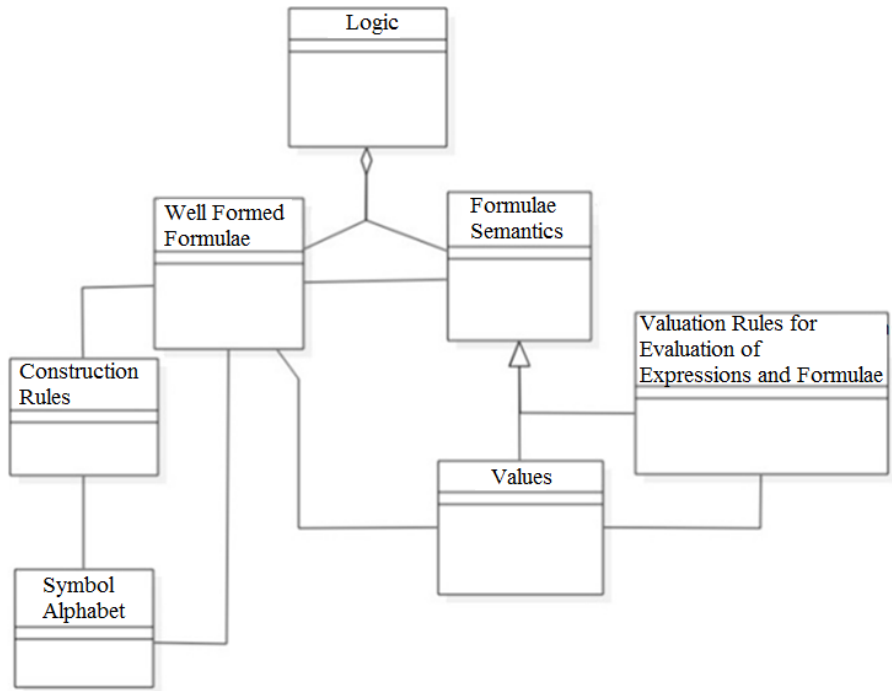

Figure 1: A meta-model for logic. (Drawn by symbolizing class diagrams in UML)

From another but similar perspective, to define logic, terms, values, the relationship between terms and values (= "Eval"), the relationship between values and values, and the relationship between terms and terms (= "Apply", for proof and formal inference) must be defined. In this approach to logic, which is close to the approach of intuitionistic logics, the concept of "value and evaluation" practically replaces "intention" and "semantics".

In intuitionistic structural logics, intention is made up of quantities and their combinations. Each logical expression is equivalent to a value. In classical logic, we have only two values, T and F. But in intuitionistic logics, a value is equal to a building of values [5]. Hence, many intuitionistic logics are considered "infinitely quantitative" logics [4] .

Each value is a construction of values, and this continues in reverse until we reach the basic values (which are basic intuitions). Hence, some sources have described the intentions of structural logic in terms such as "Bottom-Up Constructions", "Semantic Structures", "Recursive Structures", Value Lattice and the like.

Sources include Heyting Algebra-based Semantics, Kripke Structures-based Semantics, and Tarski-like Semantics for intuitionistic logics. In new sources, it seems that the common semantic model for intuitionistic logics is "Heyting-Kripke-Kolmogorov Semantic", which is also referred to as the Hitting-Kripke-Kolmogorov interpretation [4], [5], [6]. In intuitionistic logics, intention is presented through a building of values. These buildings are constructed from values based on basic values (basic intuitions, and some conventional instances of cybernetics elements i.e. objects [7], entities [7], building blocks [8], components [9], terms [10], concepts, perceptions, tasks [11] and etc.) and with the help of "rules of composition of values". There is an important correspondence between the rules of formula construction (for terms) and the rules for combining values (for intention). Each formula corresponds to a set of values. The significance of each formula (= the intention of each formula) is the corresponding values. Thus each formula, while being a formula, provides a set of values. Hence, a formula can be considered as a set of values.

In more technical terms, in these logics, the representation of intention and the representation of values are done with the help of syntactic (symbolic) structures. The intention of each term is a lattice, the leaves of which are symbols corresponding to the basic intuitions, and each node is an intermediate value, and the final value is made at the root (Figure 2). The coexistence of edges in each node represents an operation on terms (for example, the actions of the "and" logical operator). This operation affects both the term structure and the Lattice value (which means terms).

In short, in intuitionistic logics, intention is a building of values (a lattice in classical approaches or a topography in newer approaches).

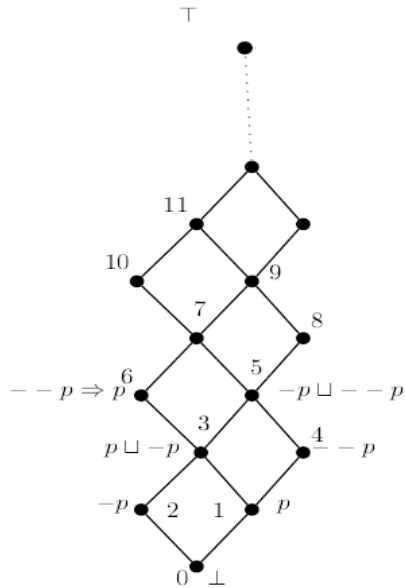

Figure 2: A lattice for values that makes up a total of a set of values. This set of values can be attributed to a term or phrase as an intention or semantics [5].

*1.1.2 Definition of intention in the proposed logic*

In this proposed logic, there is no distinction between logic and semantic model, and the terms of logic are the same as the values of the model. The syntactic structure of logic is the same as the syntactic structure of intention (which defines model values). In fact, the literal expressions and formulas of logic express exactly the values of the model. That is, we are dealing with a completely intuitionistic and completely constructive logic in which syntax is semantics.

This logic seems to be an "infinitely quantitative" logic. (Of course, many intuitionistic logics seem to be like this.)

This is an Axiomatic System that defines transactions and intentions on logical terms, representing both machine conversions on syntax and semantic conversions on semantic. Because Syntax elements are exactly the same as Semantic elements.

Semantics, according to the constructivist and intuitionistic tradition, must be a Syntactic Structure and a Symbolic Construction. In the proposed logic, this "semantic structure and structure" is the "structure and structure of the terms" of the logic itself. Instead of defining two similar structures for "term" and "intention", we present only one structure (Figure 3).

To define logic, terms, values, the relationship between terms and values, the relationship between values and values, and the relationship between terms and terms must be defined. Because this logic is structural and intuitionistic, and because the formal structure is the same as the semantic structure, all of this is provided by defining the symbolic terms logic and an axiomatic system on them.

The terms are presented in the form of well-made formulas. The intention of the terms is also expressed in the form of an Axiomatic System, which at the same time defines good construction. So "values" and "terms" and "semantic elements" are the same (in the form of a set of terms of logic). The "rules of good construction", the "rules of signification between values" and the "rules of valuation" are also the same (in the form of the Axiomatic System presented in logic).

This logic is consistent with the notion in structural and intuitionistic logics that "intention is equivalent to values, and values are recursive and symbolic constructs on basic intuitions."

In this approach, logic is the obvious facts and the intentions between them, which are included in logic in the form of terms and the intention of terms.

Every obvious semantic truth is equivalent to a value (= a term or a construction of terms) in this logic. Hence, regardless of what justification or range of intention we are talking about, we have a general logic for Semantics that can be expressed in terms of term values. This allows us to integrate modalities, semantic models and knowledge.

This interpretation of intention and Semantics (intuitionistic, value-based, and the like), in addition to intuitionistic and constructive logics, in other paradigms such as term computation (in control and cybernetics), computation with perceptions (in control and cybernetics) Combinatorial Context Grammars (in Computational Linguistics and Language Theory) and a variety of grammar approaches are also found in logical programming. There is also a similarity between this notion of intention and what exists in Functional languages and Functional programming approaches.

In short, in this approach, logic is a set of values and the signification of these values. Values which are obvious facts, and terms are set to refer to them.

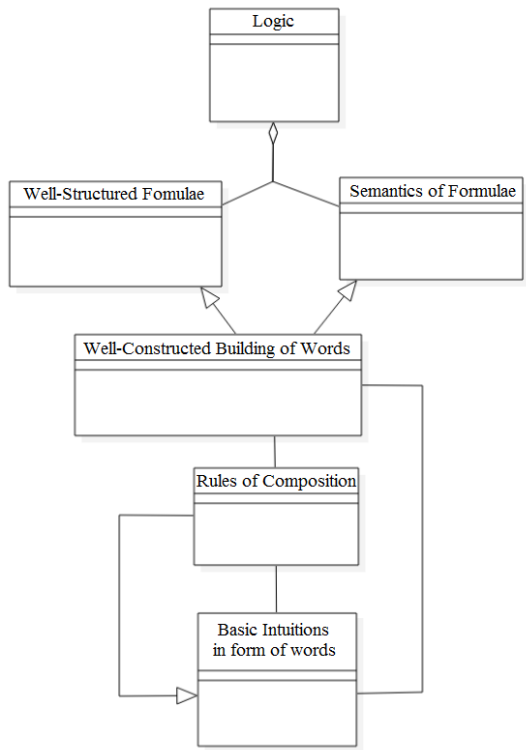

Figure 3: A meta-model for the proposed logic.

It has already been stated that an intuitionistic logic is generally defined by the presentation of two well-constructed buildings (its components and its rules) (one building for terms and one building for intention).

In the proposed logic, these two well-constructed buildings are the same. That is, logic is defined by providing a well-constructed structure (which has both the role of determining terms and how they are combined and the role of determining values and how they are combined). Because terms are only nouns for tangible values, terms are equivalent to values, and term combinations are equivalent to combinations of values. A meta-model for the proposed logic is presented in Figure 3.

*1.1.3 Definition of the proposed logic*

Processes and their surrounding intentions in business organizations are associated with a set of logical modalities and semantic models. Rules and regulations can also be defined on their own semantic modalities and models. Therefore, in order to study and solve the problem of compliance to business processes [12][13][26][27], it should be possible to somehow integrate modalities and semantic models. It has already been stated that the semantics in the Semantic Logic - that is, the proposed logic - support this.

What is meant by the integration of modalities, implications, and the integration of semantic models? That is, a composite logic can be presented that can be used to describe and solve the compliance problem. This is possible by combining modalities and semantic models, modalities and semantic models such as:

- Demands, options and choices (with operators ? And !)
- To be and not to be and existential logic (with operators $f$, $\varpi$ and !)
- Value and variable, instance, set, set membership, multiplicity and singularity (with the operators of "abstraction", "membership and inclusion", "connection", "Is-a" and "whole-component relation")
- Necessity and possibility (with O and P operators in task logic)
- Assumption, verdict and argument (with operators such as Entailment, $\Longrightarrow$, etc.)
- Numerical and arithmetic numbers and values
- Possibility

The semantic integration and connection of semantic truths of these modalities and domains together creates a semantic model composed of facts that can be expressed the intention of this model with a semantic logic.

Each of these aspects and domains has obvious facts between which there are also obvious connections. It suffices to state the terms of this logic (which are themselves obvious semantic components) along with the subject matter principles of the relationship between the terms, so that both Syntax and Semantic are defined for the proposed logic (of course, the core is independent of its domain). Further details on semantics and how to define intention have already been stated.

One of the strengths of this semantics and this logic lies in the fact that other modalities and other semantic models can be added to this logic if necessary. This is provided by adding new terms and rules to the existing set of rules. In fact, we are dealing with a logic with open semantic systems.

Then, with the help of this logic, it is possible to describe compliance issues and also, the logic and knowledge of compliance solution can be described and embedded in a process processing system. Then, by following the path in Figure 4, we can achieve levels of automation and intelligence in reviewing and resolving compliance issues.

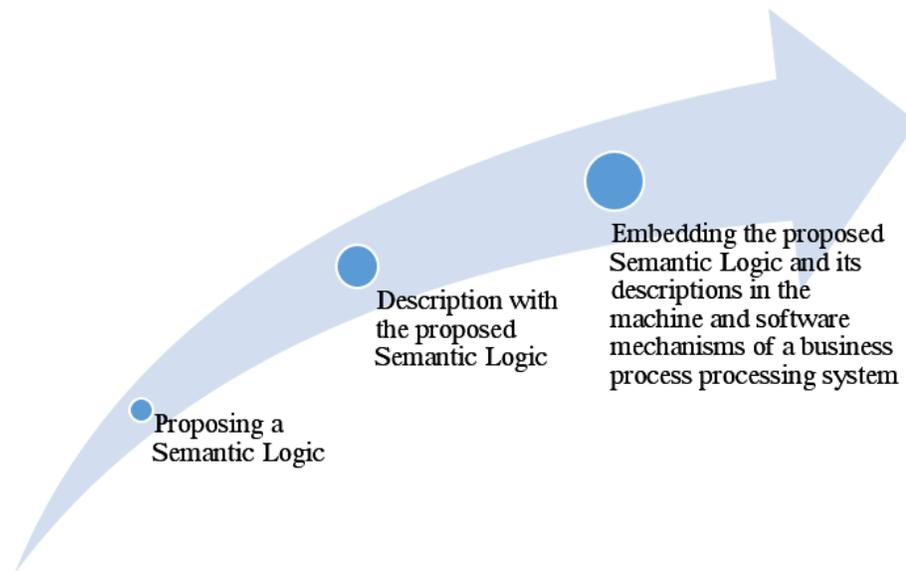

Figure 4: Path to achieve levels of automation and intelligence in solving compliance problems.

The system in which this logic is embedded must be able to receive, process and output business processes. Therefore, this logic must be embedded within a process processing system to achieve an intelligent system for reviewing and resolving compliance.

The principles of the subject (as well as the assumptions and knowledge described in each problem) help us to find signification on the combined semantic model and thereby solve the problem of compliance. For example, the principles of the following subject can be considered as part of "semantic description for the integration of semantic, existential, optional, obligatory, probabilistic, numerical, abstract and synthetic modes". (Some parentheses are out of the terms of logic and are used only for better understanding).

In these subject principles, which form the core of this logic and are considered facts independent of the domain, terms are used in exchange for obvious facts. Each of these principles [1], because of its intuitiveness, is the basis and principle:

$$
\begin{aligned}
\varphi ::=\ & \varphi \to \varphi \\
& !(\varphi)\ |\ \text{?}(\varphi) \\
& |\ \varphi \wedge \varphi\ |\ \varphi \vee \varphi\ |\ \neg\varphi \\
& |\ (\varphi)\ |\ (\varphi, \varphi)\ |\ \varphi.\varphi \\
& |\ O(\varphi)\ |\ P(\varphi) \\
& \delta\ |\ (\delta || \delta) \\
& |\ \eta > \eta\ |\ \eta = \eta\ |\ \eta < \eta\ |\ \eta <= \eta\ |\ \eta >= \eta\ | \\
& |\ \eta + \eta\ |\ \eta * \eta\ |\ \eta - \eta\ |\ \eta / \eta \\
& |\ (A{:}Z)^* \\
& |\ (0{:}9)^* \\
& |\ (0{:}9)+.\ (0{:}9)+ \\
& |\ x \qquad\qquad (x \in \mathbb{Z}) \\
& |\ x \qquad\qquad (x \in \mathbb{R}) \\
& |\ x \qquad\qquad (x \in \text{an\_Intuitive\_Set}) \\
& |\ x \in \text{an\_Intuitive\_Set} \\
& |\ x \subseteq \text{an\_Intuitive\_Set} \\
& |\ x \subset \text{an\_Intuitive\_Set} \\
& |\ \text{an\_intuitive\_Function}(x) \\
& \qquad\qquad (x \in \text{domain}(\text{an\_intuitive\_Function})) \\
\\
& |\ (\varphi\ \text{(is-a)}\ \varphi)\ |\ \varphi = \varphi\ |\ (\varphi\ \text{(is-part-of)}\ \varphi) \\
& |\ \varphi\ \text{(is-in-association-with)}\ \varphi \\
& |\ \varphi\ \text{(is-instance-of)}\ \varphi
\end{aligned}
$$

$$
\begin{aligned}
\eta ::=\ & (A{:}Z)^* \\
& |\ (0{:}9)^* \\
& |\ (0{:}9)+.\ (0{:}9)+ \\
& |\ x \\
& |\ x \\
& |\ \Pr(\delta)
\end{aligned}
$$

$$
\begin{aligned}
\delta ::=\ & (\delta * \delta) \\
& |\ (\delta \wedge \delta) \\
& |\ (A{:}Z)^* \\
& |\ \varphi
\end{aligned}
$$

*1.1.4 Domain-independent facts: Examples of the integration of modalities and semantic models*

$$?(O(A)) \land ((B \lor C) \to A) \to !(?(O(B)) \lor ?(O(C)))$$

If we demand necessity A and if B or C results in A, then we have this option: demand necessity B or necessity C.

$$!(X \lor Y) \to !(X) \lor !(Y)$$

If we have an option: "X or Y". In this case, one of the facing grins will be selectable for us[2]: "X exists" or "Y exists".

$$!(X) \land ?(X) \to X$$

If we have an option that "X exists" and if we select this option (ie we request this option), then "X exists".

$$?(P(A)) \to !(?(O(A)))$$

If we want the possibility (not impossible) of the existence of A, then there is an option to demand the necessity of the existence of A.

$$O(Pr(A*B) > alpha) \land Pr(A) \le beta \land (A \parallel B) \to$$
$$O(Pr(B) \ge alpha/beta)$$

From the necessity of greater probability of occurrence of the combination of two events A and B than the value of alpha, and from the fact that the probability of occurrence of A with beta, and from the independence of events A and B, it is resulted the necessity of more equal probability of occurrence of event B than alpha value divided by beta.

$$O(A) \land (B \to not(A)) \to not(P(B)) \land not(!(B))$$

From the necessity of A and the fact that B results in contradiction A, it follows that B is impossible and B is not the option.

$$(X \to Z) \land (Y \ (is\text{-}a) \ X) \to (Y \to Z)$$

From the fact that X denotes Z and that Y is an instance of X, it follows that Y denotes Z.

## 2 STATELESS COMPLIANCE VERIFICATION

### 2.1 Impretaive Computation / Declarative Computation

The transition function of an automata dictates an impretive computation schema for it: the concrete steps of computation are the transitions coded in this function. But if we replace this function with a declarative unit of computation (such as an axiomatic system of symbolic values and rules with a working memory), then the resulting automation has a declarative computation schema rather than an impretive one. This change never results in a change in computation power (in accordance to a turing machine) but improve the suitability of automation schema for modeling some declarative verification problems (such as compliance checking). This leads to a decrease in Kolmogorov complexity of verification program and as a result, the validation process could be done more easily.

---

[2] But which one is not clear with this information.

## 2.2 Declarative and Stateless Verification

In the method of SARV for verification, we consider these two components for a verification system:

- System Model: A lattice construction system model (rather than a state-transition).
- Semantics for Property Specification: A symbolic axiomatization of intuitions and modalities (rather than conventional semantics, such as LTL).more precisely, A set of symbols and a set of axioms define the needed semantics for property specification. An important intuition in these axioms is the System Under Study (SUS).

So we have two semantic models (defined by two semantic logic) for sytem modeling and Property Specification, respectivley. There is two important difference between them: 1) the system model intuitions may be different than Property Intuitions. 2) the System Under Study (SUS) is an important intuition in property semantics. SUS is a part of reality (or World) for an intuitionistic approach to the semantics of properties.

**Definition 1**. Each symbol which make part in a lattice construction is a symbolic value. Any complex formula that made from symbols is considered a symbol inself. Each symbolic value (based on its situaition in the constructed lattice of values) have its identity and its special meannings.

$$(\text{SUS1} = \sum_{k=0}^{n} P_i ) \textbf{ AND } (\text{SUS2} = \sum_{k=0}^{n} Q_i ) \textbf{ AND } ( \text{Ai. Ej. } P_i \text{ is-a } Q_j ) \Rightarrow \textbf{ SUS1 is-a SUS2}$$

$P_i$s and $Q_j$s are some intuitions for constructional parts. The SUS decomposition in terms of these parts should be intuitionistically intuitive. In other word, the decomposition of SUS to $P_i$s is intuitively done because its process and results are intuitive for us. So our intuition of reality (the world) makes the complex symbol of (SUS1 = $\sum_{k=0}^{n} P_i$ ) existed in our computations. Practically, there must be a piece of code or algorithm that makes and computes (SUS1 = $\sum_{k=0}^{n} P_i$ ). But in our semantic model, the (SUS1 = $\sum_{k=0}^{n} P_i$ ) could be a basic but complex symbol that directly intuisioned form the reality..

## 3 CASE STUDY: PROTOCOL COMPLIANCE CHECKING FOR SMART CITIES

Smart cities are on the way and their development, evolution, and engineering need their formal assets and languages. Each semantic space (or semantics) of involving meanings could be modeled by a separate modeling language to overcoming the complexity of the entire smart city solution.

Reo is a formal language for modeling and specification of coordination protocols [14]. An introduction of Reo syntax and behavior is mentioned in figure 5, from [15]. This language could be served as a formal modeling, verification and specification tool to engineering coordination protocols, for example see [16], [17] and [18]. Reo has a main graphical model (called Reo Circuits). Also formal semantics [19], textual notation [20], verification [21] and simulation [22] facilities are available with it and it's around ecosystem of formal coordination notations (for example see [23]).

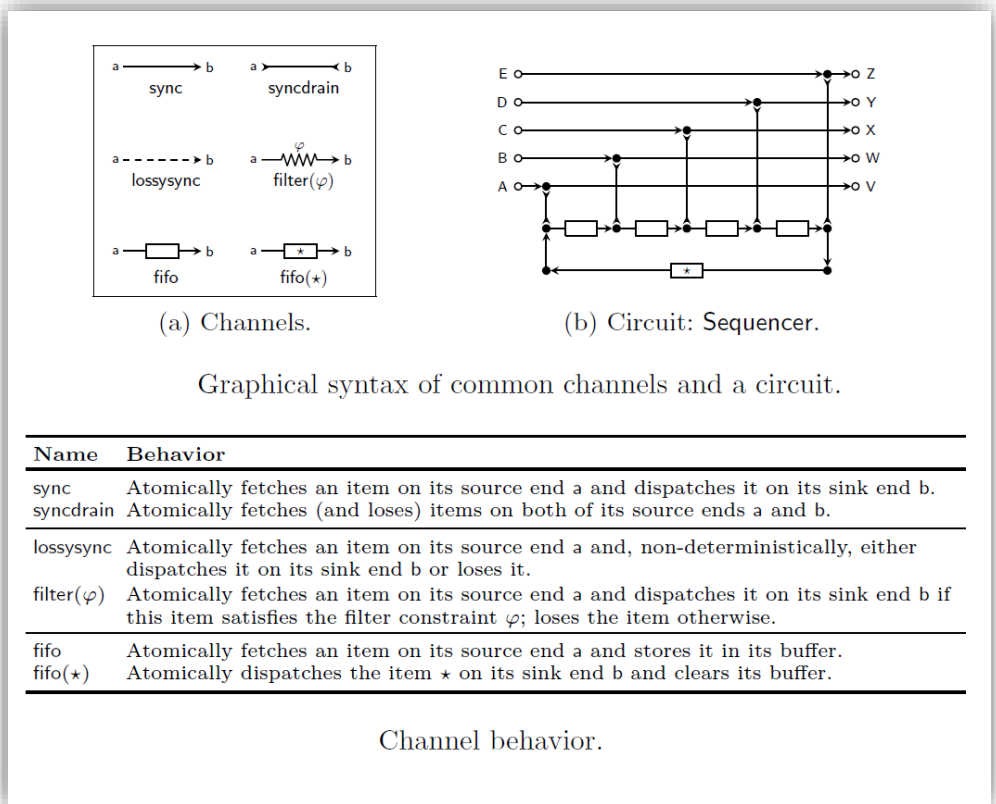

Figure 5: About Reo Syntax and Behavior, From [15].

### 3.1 Rescue Scenario Sample

Here is a very simple rescue scenario for depicting the power of Reo language to specify, model and program the coordination protocol for rescue and emergency request management, as a control module in-part of a smart city solution:

An accident situation could be detected by an approved rescue request from citizens or sensors. The call center does the approval. Then, the incoming requests are load balanced between three Emergency-Staffs. They could activate the emergency alarm after investigating the case. Then, a Police-Staff can consider the enabling of the police alarm, and simultaneously, the Firefighting-Staff can consider the enabling of the firefighting alarm.

First, a Coordination-Flow sketch of the required protocol is provided (See Figure 6). Elements of this sketch are some common reusable components for coordination-protocols, called "connectors" [14]. Then a complete implementation (in terms of a Reo circuit which is depicting the required protocol) is derived from the previous sketch (See Figure 7).

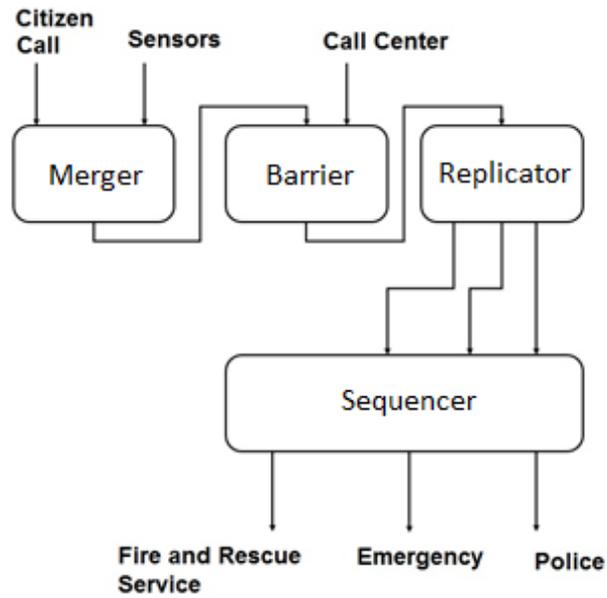

Figure 6: Coordination-Flow Sketch, by using a mesh of some common reusable components.

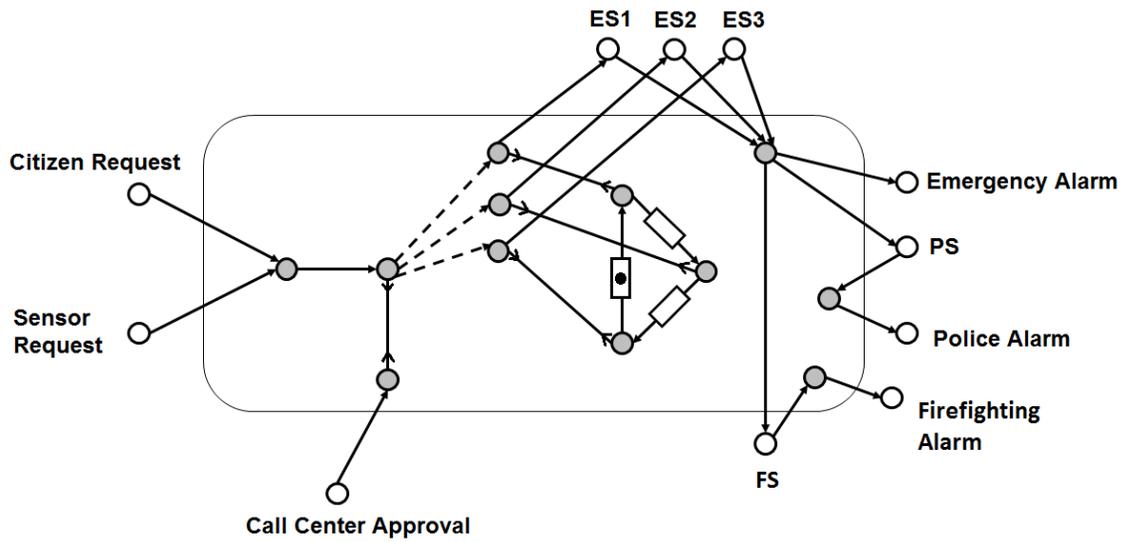

Figure 7: A Reo Circuit implementing the intended coordination protocol for handling a recue request. ES, PS and FS mean respectively Emergency Staff, Police Staff and Firefighting Staff.

## 4 SEMANTIC LOGIC PROGRAMMING FOR COMPLIANCE CHECKING

Based on Semantic Logic [24] method and KARB solution [25], we provide a logic-based programming to implement an instance of Compliance Checking for the protocol of our example (based on related semantic domains, figure 8). Here is the code and the yielding symbolic calculations (figure 9):

```
Protocol
1.      AmbulanceRequest >> FireRequest >> PoliceRequest
2.      PoliceRequest => P(HelicopterMisson)
3.      FireRequest => P(HelicopterMission)
4.      AmbulanceRequest => P(HelicopterMission)
5.      HelicopterMission => BudgetConsuming

Compliance Rules
6.      Forbidden((Very)BudgetConsuming)

Obligation Semantics
7.      Forbidden(A) AND P(A) => Warning(P(A))
8.      Forbidden(A) AND A => Failure(A)
9.      Warning(P(A)) AND DoubleCheck(P(A)) => Resolved(Warning(P(A)))

Counting Semantics
10.     A AND (I)A => (I+1)A
11.     A =>(1)A
12.     (I)A AND I>2 => P((Very)A)

Deontic Semantics
13.     (A=>B) AND (P(A)) => P(B)
14.     (Very)P(A) => P((Very)A)
15.     P(P(A)) => P(A)
```

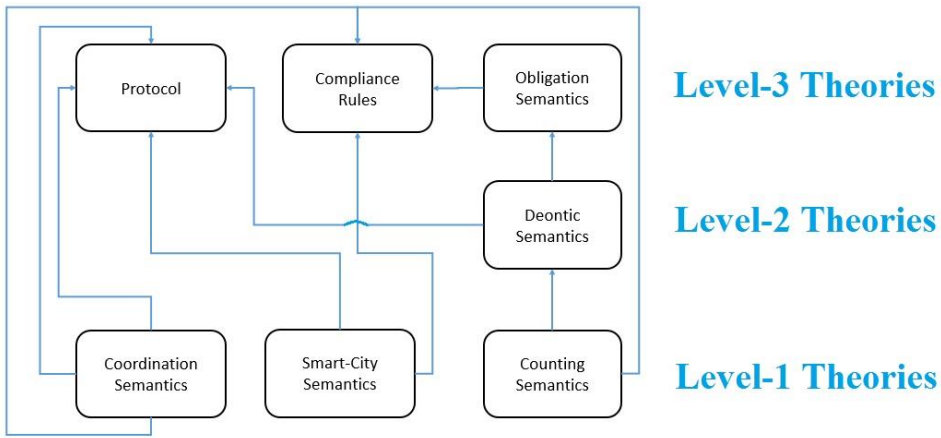

Figure 8: The Relation Between the involving Semantic Models.

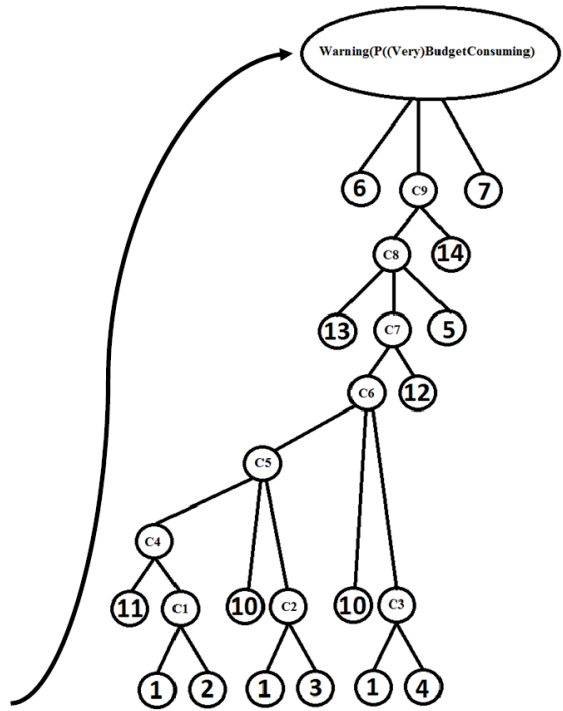

Figure 9: The Symbol-Generation Tree (in accordance to the entailment process).

## 5   AUTOMATION OF SARV BY SHAMS

On top of a java-based Semantic-Logic programming framework (named SHAMS ), we could program, compute and generate a symbolic lattice of a given semantic logic. An example of an automatically-grown symbolic lattice (figure 10) is provided, which is generated for a given semantic logic (figure 11) on this input value:

```
"How(Excitement(Ability(See(Unseen))))"
```

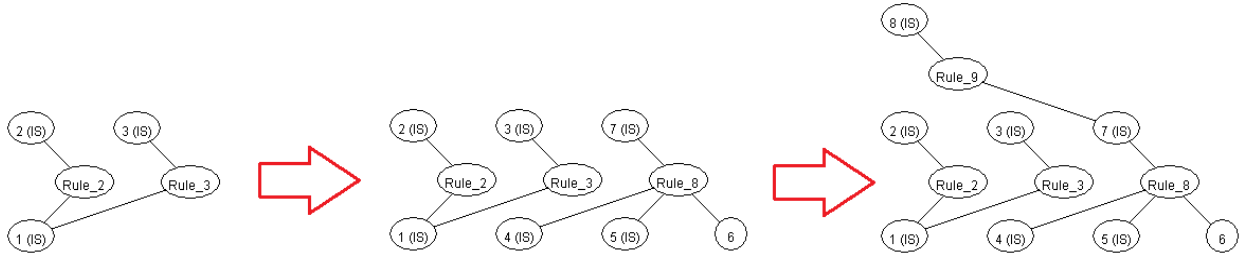

Figure 10: An automatically-grown symbolic lattice

```
1    ============Input Rules =========================================
2    Question ( Ability ( A ) )    ======>  Abduction ( P ( Not ( A ) ) )
3    How ( Excitement ( Ability ( See ( Unseen ) ) ) )  Is-In  Beginning
4    A ( B )  AND  Wonder ( B )  AND  A Is-A Positive_Sense    ======>  Wonder ( A ( B ) )
5    Engagement  AND  Excitement    ======>  Propagation ( Engagement )
6    A ( B ) <> B    ======>  Wonder ( A ( B ) )
7    Engagement    ======>  Attention_Policy ( Engagement )
8    Ability Is-A Positive_Sense
9    Question ( A )  Is-In  Beginning    ======>  Engagement ( A )
10   See <> Unseen
11   Wonder ( A )  AND  A  Is-In  Beginning    ======>  Attention_Policy  AND  Excitement ( Wonder ( A ) )
12   How Is-A Question
13   How ( Excitement ( Ability ( See ( Unseen ) ) ) )
14   Excitement    ======>  Attention_Policy ( Excitement )
15   How Is-A Positive_Sense
16   Engagement  AND  P ( Not ( Excitement ) )    ======>  P ( Not ( Promotion ) )  AND  P ( Not ( Excitement ) )
```

Figure 11: A given symbolic logic

## 6   SARV FOR QUALITY COMPLIANCE: DD-KARB METHOD

Rather than proof and logical evaluation, benchmarking can be used for compliance assessment. Naturally, a set of benchmarks can shape an applied solution to compliance assessment. The KARB solution system is based on this approach [25], i.e. preventing compliance anomalies through SARV rule-based benchmarking. In fact, SARV rule-based benchmarking means evaluating an under-compliance system with its symbolic specification and using a set of symbolic rules (on the behalf of the semantic logic of evaluation). A case study was conducted to demonstrate and analyze the KARB solution. The IR-QUMA study (Iranian Survey on Quality in Messenger Apps) was conducted to evaluate the quality of some messenger applications. According to the evaluation results, the hybrid DD-KARB method (with a combination of semantics-awareness and data-drivenness) is more effective than famous machine learning methods [28] (See Figure 12) and can compute a somehow good estimation for the messenger application user quality scores. Therefore, DD-KARB can be considered a method for quality benchmarking in this technical context. Details about DD-KARB method is available from [25].

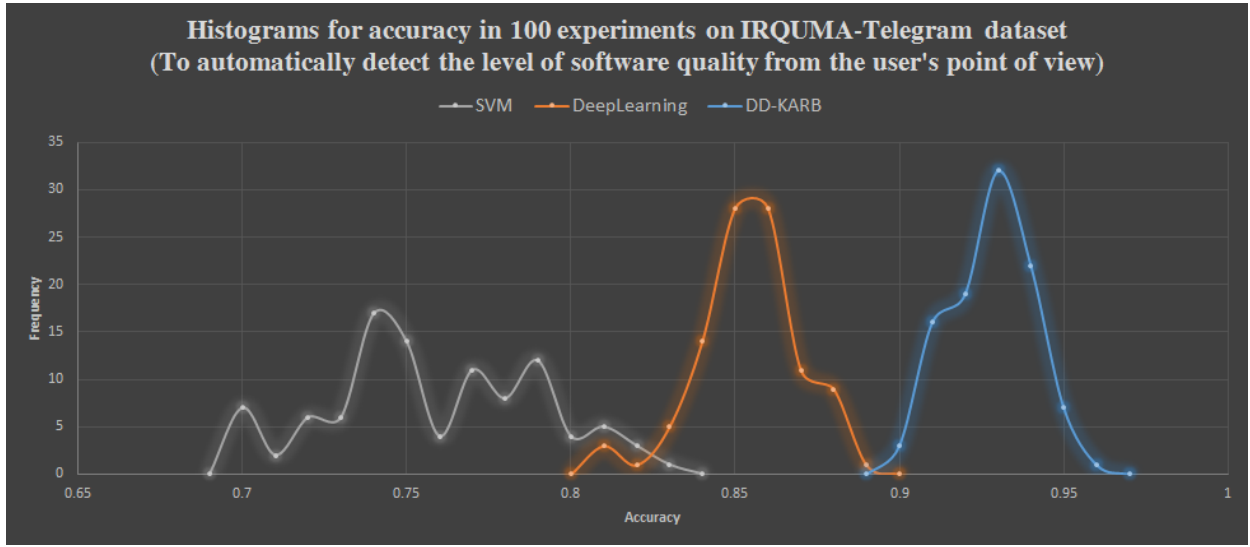

Figure 12: DD-KARB outperforms SVM and Deep Learning on IR-QUMA-Telegram dataset, in automatic detection of complete quality satisfaction (detection of the fifth Likert level from four other levels). See [28] for details of data experiments settings and results.

Based on intuitionistic approach, an underlying logic for computation (for the problem and the system under study) is prepared. Then based on this logic, which is named "semantic logic", we construct a semantic infrastructure for the problem, system and solution, called the qualifier. Then we gather a gold benchmark dataset for learning. The error function of computed attributes (based on the qualifier) is considered as a minimizing fitness function. By utilizing several optimization algorithms or heuristics, we could find an optimum for this fitness function. Then we have a learned qualifier function that is result of compliance solving of intuitions (i.e. semantics) to data. This hybrid approach uses intuitionistic logic, semantics and data together for machine learning of solutions for problems.

# 7 CONCLUSION

With the help of semantic logic, intelligent systems can resolve compliance. In this way, the logic of describing the system under compliance and the logic of resolving compliance are expressed with the help of formal semantics in semantic logic, and then the resulting semantic logic is embedded in a processing context and made use of (For example, with the help of the logical programming paradigm - for example with prolog - it is possible to embed the logic of intentions (i.e. Semantic Logic) in the computational system and resolve the compliance based on it).

Coordination protocols are among key elements of complex systems. Smart cities, as complex systems, need their coordination protocols. We introduced a solution for programming and compliance checking of coordination protocols for smart cities. Because of easiness of translation for Reo language semantics to semantic logic, we could simply compliance-check the protocols against our proposed conditions.

The results of machine learning data experiments on IR-QUMA dataset suggest that the SARV-based DD-KARB can be considered a method for quality benchmarking.

**ACKNOWLEDGMENTS**

We would like to express our gratitude to the members of Sharif DiSysLab Laboratory, especially Mr. Farhadi (PhD candidate).